\documentclass[conference,compsoc]{IEEEtran}
\makeatletter
\newcommand{\linebreakand}{%
  \end{@IEEEauthorhalign}
  \hfill\mbox{}\par
  \mbox{}\hfill\begin{@IEEEauthorhalign}
}
\makeatother

\usepackage{graphbox}
\usepackage{comment}

\usepackage{ulem}

\ifCLASSOPTIONcompsoc
  \usepackage[nocompress]{cite}
\else
  \usepackage{cite}
\fi
\ifCLASSINFOpdf
\else
\fi

\usepackage{color}

\usepackage{enumitem}

\usepackage{etoolbox}
\makeatletter
\patchcmd{\@verbatim}
  {\verbatim@font}
  {\verbatim@font\footnotesize}
  {}{}
\makeatother

\usepackage[utf8]{inputenc}
\usepackage{dirtytalk}

\newcommand{\squeezeupa}{\vspace{-.5cm}}

\usepackage{fancyvrb}
\usepackage{pgffor}
\usepackage{multicol}

\begin{document}
%
\title{Development of Authenticated Clients and
 Applications for ICICLE CI Services\\
\small{Final Report for the REHS Program, June - August, 2022}
}

\author{\IEEEauthorblockN{Sahil Samar }
\IEEEauthorblockA{Del Norte High School, \\
San Diego, CA, USA\\
Email: sahilsamar031@gmail.com}
\and
\IEEEauthorblockN{Mia Chen}
\IEEEauthorblockA{Westview High School,\\ 
San Diego, CA, USA\\
Email: mialunachen@gmail.com}
\and
\IEEEauthorblockN{Jack Karpinski}
\IEEEauthorblockA{ San Diego High School, \\
San Diego, CA, USA \\
Email: jackadoo4@gmail.com}
\and
\IEEEauthorblockN{Michael Ray}
\IEEEauthorblockA{JSerra Catholic High School, \\
San Juan Capistrano, CA, USA \\
Email: michael.ray@jserra.org}
\and
\IEEEauthorblockN{Archita Sarin}
\IEEEauthorblockA{ Mission San Jose High School, \\
Fremont, CA, USA \\
Email: archita.sarin@gmail.com}

\and
\IEEEauthorblockN{Christian Garcia (mentor) }
\IEEEauthorblockA{Texas Advanced Computing Center, \\
Austin, TX, USA \\
Email: cgarcia@tacc.utexas.edu}
\and
\IEEEauthorblockN{Matthew Lange (mentor)}
\IEEEauthorblockA{ IC-FOODS, \\ 
Davis, CA, USA \\
Email: matthew@icfoods.org}
\and
\IEEEauthorblockN{Joe Stubbs (mentor)}
\IEEEauthorblockA{Texas Advanced Computing Center, \\
Austin, TX, USA \\
Email: jstubbs@tacc.utexas.edu}

\and
\IEEEauthorblockN{Mary Thomas (mentor) }
\IEEEauthorblockA{San Diego Supercomputer Center\\
La Jolla, CA, USA \\
Email: mpthomas@ucsd.edu}}

\maketitle
\thispagestyle{plain}

\begin{abstract}
 The Artificial Intelligence (AI) institute for Intelligent Cyberinfrastructure with Computational Learning in the Environment (ICICLE) is funded by the NSF to build the next generation of Cyberinfrastructure to render AI more accessible to everyone and drive its further democratization in the larger society.
We describe our efforts to develop Jupyter Notebooks and Python  command line clients that would access these ICICLE resources and services using ICICLE authentication mechanisms.
To connect our clients, we used Tapis, which is a framework that supports computational research to enable scientists to access, utilize, and manage multi-institution resources and services.
We used Neo4j to organize data into a knowledge graph (KG). We then hosted the KG on a Tapis Pod, which offers persistent data storage with a template made specifically for Neo4j KGs. 
In order to demonstrate the capabilities of our software, we developed several clients: Jupyter notebooks authentication, Neural Networks (NN) notebook, and command line applications that provide a convenient frontend to the Tapis API. 
In addition, we developed a data processing notebook that can manipulate KGs on the Tapis servers, including creations of a KG,  data upload and modification. 
In this report we present the software architecture, design and approach, the successfulness of our client software, and future work.

\end{abstract}


%
\IEEEpeerreviewmaketitle

\pagestyle{plain}


\section{Introduction}


\noindent The \textit{AI institute for Intelligent Cyberinfrastructure with Computational Learning in the Environment} (ICICLE) program aims to develop intelligent cyberinfrastructure with transparent and high-performance execution on diverse and heterogeneous environments, as well as to advance plug-and-play AI that is easy to use by scientists across a wide range of domains, promoting the democratization of AI.\cite{icicle-proj}\cite{Parashar2022}
The deep AI infrastructure that ICICLE plans to utilize to sift through data relies on knowledge graphs (KGs) in which information is stored in a graph database that uses a graph-structured data model to represent a network of entities and the relationships between them.\cite{kg-wiki} 

Finding a way to make KGs easily accessible and functional on high performance computing (HPC) systems is an important step in helping to democratize HPC. Thus a large focus of this project was contributing to the body of knowledge needed for hosting live, dynamic, and interactive services that interface with HPC systems hosting KGs for ICICLE based resources and services. 
 
\begin{figure*}[t]
\centering
\includegraphics[width=.9\linewidth]{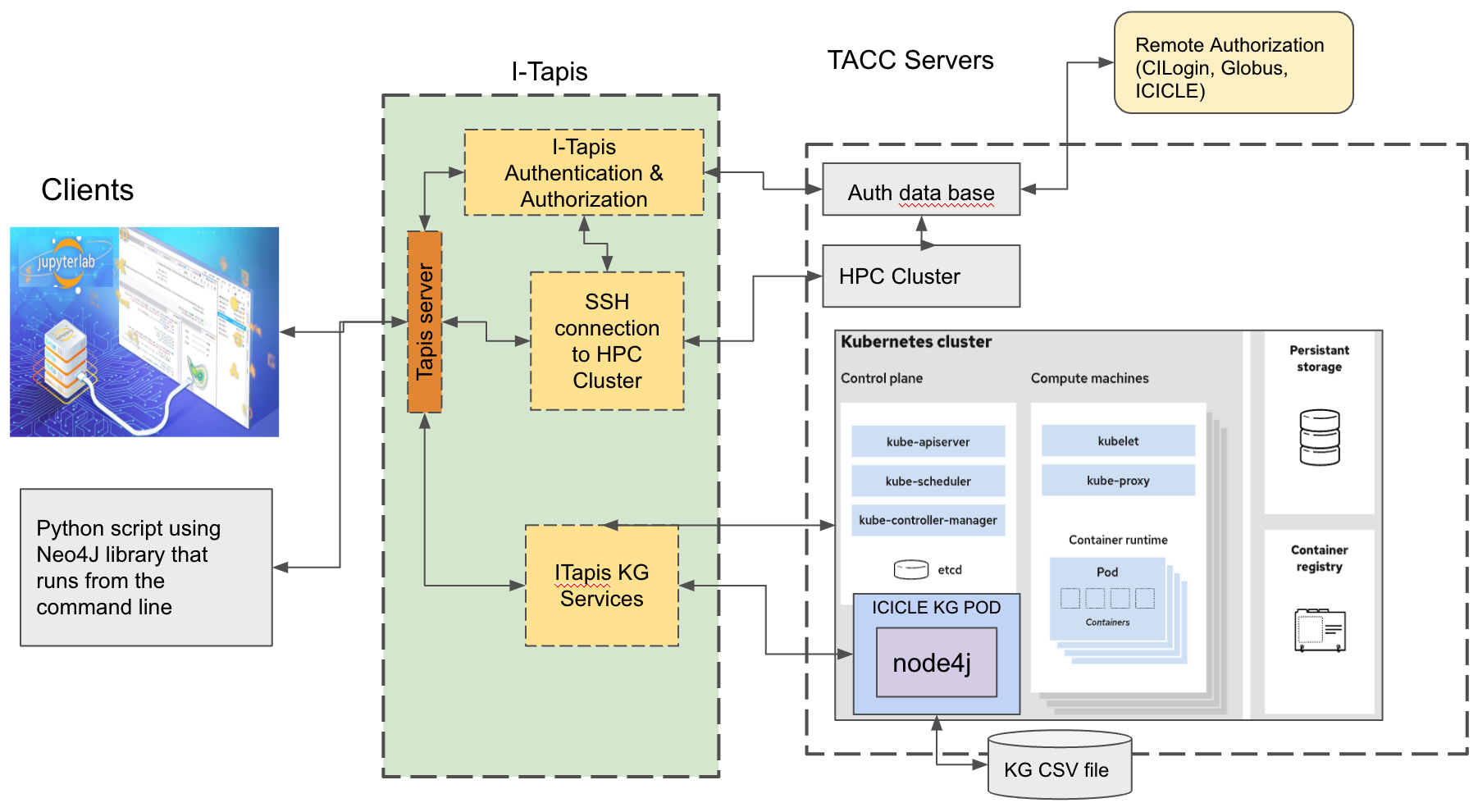}
\caption{Hello ICICLE 3-tier middleware architecture, consisting of client, middleware and backend resources.} 
\label{fig:hello-icicle-arch}
\end{figure*}

This report describes the results and lessons learned from the project we worked on as part of the San Diego Supercomputer Center (SDSC) \cite{sdsc-rehs} 2022 Research Experience for High School Students (REHS)  program.\cite{rehs-proj-2022},  

 During the course of the project, we learned about many technologies, in the following core areas: (1) Programming and Development - Bash scripting, Docker containers, Jupyter Notebooks, JupyterLab, Anaconda Environments, Machine learning, Neo4j Cypher Query Language, CLI Development, Github; (2) Tapis –  instantiating Tapis Pods on remote HPC systems, exercising the TapisAPI in notebooks/python scripts; using Tapis to remotely run singularity/docker containers, using Tapis as an alternative to SSH for file transfer; and (3) HPC Systems – connecting to and running jobs on Expanse and Stampede, How to use the Slurm batch scheduler, How to navigate the Expanse storage system.

In the following sections, we describe the architecture and components of the system as follows: Section~\ref{sec:hello-icicle-arch} describes the architecture of the Hello ICICLE system and the various components of the sofware stack; Section~\ref{sec:methodology} covers the methodology we used for creating Knowledge Graphs, developing Jupyter Notebook and Python Clients, and testing applications such as Neural Networks and Machine Learning with KGs; Section~\ref{sec:results}  summarizes the results of our project; finally, Section~\ref{sec:conclusions} provides a conclusion as well as an outline of further research to be done with the ICICLE project.

\section{Architecture of the Hello ICICLE System}
\label{sec:hello-icicle-arch}
\noindent There are many ways to access HPC systems, but they are generally difficult and require a specialized understanding of computer science and system architecture. Many scientists and researchers who want to analyze large data sets do not have this expertise, and cannot utilize advanced computing resources to their fullest. As a result, there is a high demand for simpler, more user friendly interfaces with HPC systems. Many of these architectures are based on a 3-tier middleware architecture \cite{Thomas2010}, which consist of a client, middleware and backend resources. The middleware layer often includes the Web app frameworks, and local databases and services.
To meet this demand, we developed a system we initially referred to as \textit{Hello ICICLE} that allows clients to create, edit, and query remote ICICLE services (see Figure~\ref{fig:hello-icicle-arch}). The system supports Jupyter Notebook and command line interface (CLI) clients, that use the Tapis Framework \cite{tapis-proj-doc} to authenticate and run jobs on remote services and to access Neo4J generated Knowledge graphs \cite{neo4j-proj}.  The various components used in the Hello ICICLE system, what we learned about them, and how they are used are described in the sections below.

\subsection{Hello ICICLE Clients}
\label{sec:hello-icicle-clients}
\begin{figure*}[t]
\centering
\includegraphics[width=.8\linewidth]{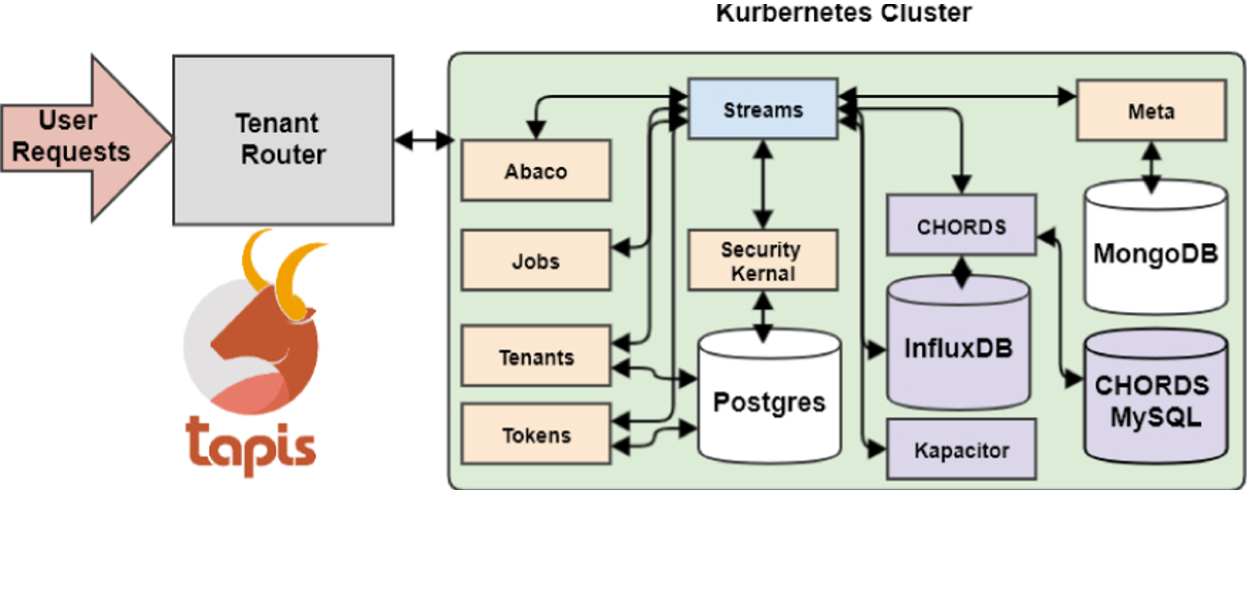}
\squeezeupa
\squeezeupa
\squeezeupa
\caption{Figure shows how user requests (e.g. Jupyter Notebook) are routed to the appropriate Pod service hosted in the Kubernetes cluster(green). \cite{Cleveland2020}} 
\label{fig:tapis-workflow}
\end{figure*}

\noindent We designed two types of clients that can be run on the users local laptop, workstation or HPC system: Jupyter Notebooks and command line Python apps (CLI). Jupyter Notebooks are popular Web based applications that compartmentalize and simplify authentication and interfacing to jobs and services hosted on these HPC systems.\cite{Milligan2018, Stubbs2020} We initially ran and tested the notebooks provided by the Tapis project. To create an ICICLE relevant test case, we focused on the deployment and use of Neo4j knowledge graph databases using pods hosted on Kubernetes clusters that are hosted via Tapis and run on TACC’s \textit{Cyclone} supercomputer. Figure ~\ref{fig:tapis-workflow} shows how user requests (e.g. the Jupyter Notebook) are routed to the appropriate Pod or service hosted in the Kubernetes cluster(green).

Many HPC systems users know how to run programs using command line interface (CLI) applications. Consequently, we developed a set of CLI applications that emulate the Jupyter Notebook KG functions. Results and outcomes of our clients are discussed in the Results section ($\S$~\ref{sec:results}).

\subsection{Tapis Computing Platform}
\label{sec:tapis-arch}
\noindent Tapis authentication is a multi-layered operation that constitutes the most difficult portion of this process. Figure ~\ref{fig:tapis-seq-diagr}
shows the backend architecture of the Tapis system, as well as the Tapis API.\cite{Cleveland2020} In our project, we heavily leveraged the use of the Tapis API in order to create a Neo4j template database hosted on Tapis systems. User requests are routed through the Tenant Router, and authentication is done using Tokens and the Security Kernel; then, a user’s request for a job is able to go through. The workflow for the end user is much simpler as these layers are handled by the Tapis API.

\subsubsection{Tapis Authentication and Authorization}
\label{sec:tapis-auth}
\noindent Tapis has provided a tutorial called “Hello, Tapis”  in order to introduce researchers to the Tapis systems.\cite{icicle-proj} We have compiled a Jupyter Notebook \cite{rehs-proj-2022} in which we simplified and elaborated parts of this tutorial and annotated each step with clear instructions.

For authentication and authorization via Tapis, we leveraged TACC accounts. The Jupyter Notebook starts with a request for input of the TACC username and password which will then be stored as variables in the notebook. Then, Tapis is imported and a python Tapis client is created for the user. After a call to the Tokens API, the Notebook prints out a string that includes a token that the user needs to set to the JWT variable in the shell using the command:  
\begin{verbatim}
export JWT=<access token starting with ey…> 
\end{verbatim}
Then the user can run the command: 
\begin{verbatim}
t.authenticator.get_userinfo()$
\end{verbatim}
 
\noindent This command should return the info that is part of the user’s TACC account. The next step is to access a new system. To do this you need to create a system variable which in our Notebook is used to access Stampede2 by someone with an allocation on it. To run it, edit cell 5 in the Notebook by replacing \textless stampede\_username\textgreater in the code with your stampede username. Then after running a createSystem command, you have connected to the system. and there are some commands to check if you have connected correctly.

\subsubsection{Using Tapis Pods}
\label{sec:tapis-pods}
\noindent 

There are a multitude of possibilities of what can be done once authenticated. Accessing a TACC machine like Stampede or Jetstream requires that HPC system to be registered with the TACC account. We cover this process in our “Tapis-Notebook” notebook. Most of our project, however, was focused on Tapis Pods. Tapis Pods are a Tapis service which allow for easy deployment and use of databases. They are able to import and export data from live databases. Tapis has an easy to use template to create a pod that will host Neo4J knowledge graphs. 

Each Tapis pod hosts one knowledge graph. These are stored on TACC systems using Kubernetes. The creation of these Pods is simplified through the Tapipy library. It is simply done through this method of the tapis object, “t”: 
\begin{verbatim}
t.pods.create_pod(pod_id=pod_id, 
    pod_template=pod_template,
    description=pod_description)
\end{verbatim}

\noindent Each of the parameters should be defined prior to using them. The \say{pod\_id} is a string that will be used to reference the pod. The “pod\_template” should describe what is going into the pod. The pod\_template that we used was a neo4j template, which is specified simply through assigning the parameter a string value of \say{neo4j}. The third parameter provides some information about the pod.

\subsubsection{Tapis Generated User Credentials}
\label{sec:tapis-gen-creds}
\noindent The next section describes the steps needed to create user credentials, which are used to access the pods. For example, each of the pods (described in Section~\ref{sec:tapis-pods}) has a specific set of permissions, and one can have the same permissions as the owner with the \say{ADMIN} level. These permissions can be set using our Tapis console application or can be done through a few simple commands, which are also covered in the \say{tapis\_auth} notebook. When pods are created, they have preset credentials associated with them. These credentials can easily be accessed from a TACC user’s python Tapis object through the following code: 
\vspace{1.5pt}
\begin{verbatim}
username,password=t.pods.get_pod_credentials
(pod_id = pod_id).user_username,t.pods.
        get_pod_credentials(pod_id=pod_id)
        .user_password
\end{verbatim}

\noindent where \say{t} is the Tapis object. It is important to not print out these passwords, as it is currently a security threat since once logged in, the credentials can be accessed, but those credentials can be used by members without any other access to the pod. The current best practice is to do as done in the code snippet, where the credentials are stored in aliases that are referenced when the user needs to login to a pod. Currently, the best supported and easiest way to create a template using pods is the Neo4j template. Once a Neo4j pod is created and the permissions have been set, the authentication flow is as follows:

\begin{enumerate} 
  \item Using python, create a Tapis object.
  \item Login to TACC.
  \item Get the ID of the pod you want to access.
  \item Get the credentials of that pod.
  \item Pass in those credentials to a Neo4j graph object.
  \item Use the Py2Neo client package to generate the graph.
\end{enumerate}

\noindent An example for generating the graph is shown below:
\begin{verbatim}
graph = Graph(f"bolt+ssc://{pod_id}.
pods.icicle.develop.tapis.io:443", 
auth=(username, password), secure=True, verify=True)
\end{verbatim}

\noindent In the above example, “icicle.develop.tapis.io” is the base url that was used to create the pod. The base url is specified in the creation of the Tapis object, like so: 
\begin{verbatim}t = Tapis(base\_url =
"https://icicle.develop.tapis.io", 
username = username, password = getpass('password'))
\end{verbatim}

After the Pod is completely registered and all of the above steps have been followed, authentication can be largely streamlined using the resources that we have made such as our console applications and Jupyter Notebooks. All of these resources essentially just require a TACC login (as of now), and handle all of the further authentication described in this section in the backend.

\subsection{Neo4j Database Service}
\label{sec:neo4j}
\noindent Neo4j is a graph database service which allows users to create complex visualizations of data by using node edges and relationships instead of columns and rows to arrange data. Each node holds data, and connects to one or more nodes through relationships. Both nodes and relationships can possess unique properties to describe the data. The Neo4j article, ``From Graph to Knowledge Graph: How a Graph Becomes a Knowledge Graph", provides a simple definition of a KG to be ``an interconnected dataset enriched with semantics so we can reason about the underlying data and use it confidently for complex decision-making."\cite{neo4j-defn-kg} 

Neo4j allows for flexibility in storing relationship driven data. Where relationships or classifications play a significant role, Neo4j proves much simpler and more efficient than other services, such as SQL. Throughout our project, a main testing focus was using python and Tapis to manage Neo4j knowledge graphs remotely on Stampede.

\section{Methodology}
\label{sec:methodology}
\noindent In order to test the architecture, we had to develop software and clients to access the different services that are part of the overall architecture (see Figure~\ref{fig:hello-icicle-arch}). These are described below.

\subsection{Client Authentication Processes} 
\label{sec:client-auth}
\begin{figure}[t]
\includegraphics[align=b,width=.96\linewidth]{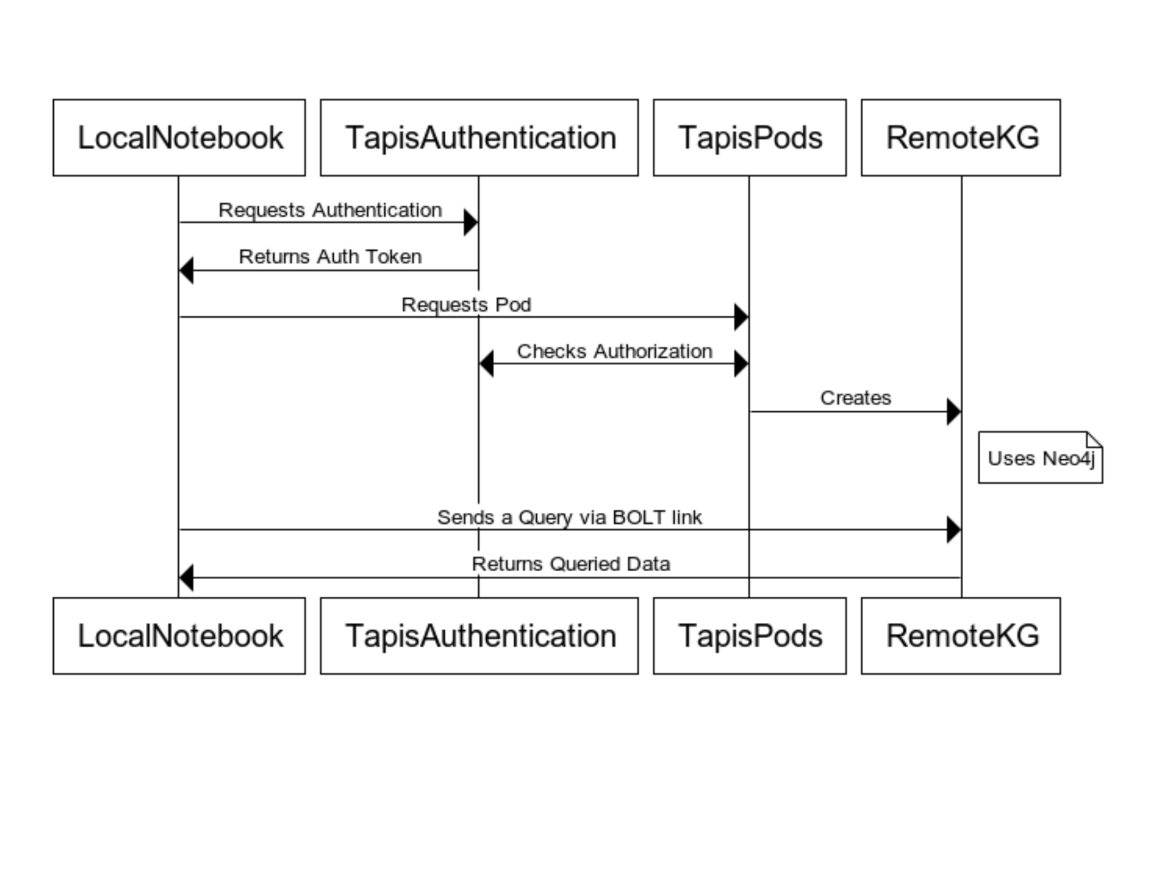}
\squeezeupa
\squeezeupa
\squeezeupa
\caption{Sequence diagram showing access steps needed to authenticate the Jupyter Notebook client to the Tapis authentication system} 
\label{fig:tapis-seq-diagr}
\end{figure}

\noindent The first step was the authentication. That only required a TACC username and password. However, there were multiple steps to follow to get the knowledge graphs to run on the Tapis Pods. In order to understand the authentication process that we went through to access our knowledge graphs, we created sequence diagrams of the Tapis architecture. These diagrams contained the steps that connected our laptop to the pods where the notebooks were run. Figure~\ref{fig:tapis-seq-diagr} shows the  steps needed to authenticate the Jupyter Notebook client to the Tapis authentication system.

Once Tapis authenticates the TACC username and password from the client notebook, it returns a token that the user uses to request a pod. The software checks the authentication and if everything properly executes, it creates a pod on which the knowledge graph is remotely hosted. The KG runs on the pod using Neo4J. To query the data, the computer sends a query through the BOLT link and the remote KG returns the output.

\subsection{Creating Knowledge Graphs }
\label{sec:create-kgs}
\noindent We initially used a Neo4j service to get acquainted with creating knowledge graphs. Using cypher, we each created our own knowledge graphs and then worked together on a larger set of data to efficiently host it on ICICLE systems. It was integral that we had a larger set of data in order to confirm that hosting the KG would be easy for scientists looking to just test their data. We found a way to write, query, and edit the KGs with Jupyter Notebooks using a Neo4j python package called ``py2neo''. We then automated the creation of knowledge graphs from .csv files via the LOAD CSV cypher command in a Jupyter Notebook. In the final product of our software, we exclusively used the Tapis pods service to create our KGs.  After authentication on TACC, all that was left is to run the KGs. We created Jupyter Notebook clients to efficiently host remote KGs.  

\subsection{Accessing and Hosting Knowledge Graphs}
\label{sec:neo4j-hosting}
\noindent As part of our project, we investigated several mechanisms for hosting KGs for public consumption. These included locally hosted graphs on our laptops, using the Neo4J browser and desktop apps, and the AuraDB and Bolt services. As we tested these methods, we determined that Kubernetes Pods (hosted on Tapis) were the optimal solution for our experiments. These methods are described below.

\subsubsection{Local Host}
\label{sec:neo4j-local}
\noindent Our project primarily used Neo4j’s localhost service for testing purposes, which was useful given its simplicity and speed. While localhost services cannot operate on remote computers or be shared between team members, for much of the project it allowed us to quickly test cypher queries and programs without having to worry about networking.

\subsubsection{Neo4j Browser and Desktop Apps}
\label{sec:neo4j-apps}
\noindent The Neo4j browser can be used for lightweight visualization of knowledge graphs. The Neo4j browser can be used to connect to a Tapis Pod with a Neo4j template. The credentials for authentication for this connection can be found using the same methods as described in the previous section. The Neo4j browser can be accessed through the Neo4j Desktop app. From there, a new connection can be created; the checkbox for using an encrypted connection must be ticked here to authenticate. Then, the credentials for the Pod simply have to be entered and the Neo4j Browser can open the project. Cypher queries can be visualized from here. The main drawback was that the browser and desktop apps were unable to host the KGs themselves.

\subsubsection{AuraDB Hosting Service}
\label{sec:neo4j-auradb}
\noindent AuraDB is Neo4j’s own cloud service for hosting knowledge graphs. While it is useful for people who don't want to go to the trouble of manually setting up a bolt service, it isn't persistent, and not easily scalable due to the additional costs of hosting more than one knowledge graph. As such, for large scale or multi-project data hosting, AuraDB is not the ideal option. As such, while we initially used AuraDB for our database hosting, we quickly phased it out for other methods.

\subsubsection{Bolt Portforwarding}
\label{sec:neo4j-bolt}
\noindent After trying and realizing the limitations of cloud based solutions, Neo4j’s Bolt service was our next option. We used our own laptops as servers, and exposed a Bolt URL for public access. While it did provide a cost free way to host our knowledge graphs, setup was difficult and unreliable. In the few instances we managed to set up bolt services, they were unstable (dropping connections, service errors), and not easily scalable, since each database had to be manually configured (if no other software is used). This made Bolt unsuitable for our intended use.

\subsubsection{Tapis Pods}
\label{sec:neo4j-tapis-pods}
\noindent Tapis pods were our final solution, and fixed most of the aforementioned issues with other graph hosting services. As shown in Section II, after authenticating with  the Tapis service, the user can create, or access a graph database in a Kubernetes cluster. Since the entire database creation and querying backend was streamlined by TACC, it takes minimal python code and minutes to accomplish the same as with  hours configuring bolt. Thanks to its user friendliness and reliability, Tapis pods became the primary hosting service for our project.

\subsection{Developing the Jupyter Notebook Clients }
\noindent A key goal of our project was to create Jupyter Notebook Clients that allowed anyone to easily access Stampede2 and other Tapis supported HPC’s. We wanted the Notebooks to be able to run from a local computer (laptop, workstation) and access remote HPC systems, so we used the Tapis Authentication system. Some of these Notebooks are a slight variation on the “Hello Tapis” tutorial.\cite{hello-tapis} Others allow people to access and visualize KGs using the py2neo, neo4j, and neo4jupyter python modules. These notebooks can be found in the project GitHub repository.\cite{rehs22-github-icicle}

To create these Notebooks as a team, we hosted them on our Github repository.\cite{rehs22-github-icicle} We used Visual Studio Code and Github Desktop to write and push edits to our Notebooks. For testing, we ran our Jupyter Notebooks using the browser, Visual Studio Code, and HPC environments.

All of this effort was in order to make HPC resources more accessible for computational scientists without needing to go through all the training that is currently required to use them.

\subsection{Developing Python Clients }
\noindent Not only did we create Jupyter Notebooks but we also made python clients that can be run from the command line. These clients access the KGs hosted on the TACC system using a few simple inputs. It does require your TACC username and passcode for authentication. It shows you all the pods that you have access to and asks which you'd like to use. It then allows you to easily access, create, delete, change permissions, get information, and query the knowledge graph without needing to know Cypher (Neo4j’s query language).

\section{Results}
\label{sec:results}

\noindent In the next sections we describe the projects that were developed by this project. Some of the projects were done in collaboration by the entire group, while some were done by individuals; but we were always helping each other with ideas and solutions. All of the software created by the students can be found in the project GitHub repository.\cite{rehs22-github-icicle}. In addition, some of our work (ICIConsole and TapisCL-ICICLE) will be included in the first ICICLE Software Components Catalog (VC3), which released around April 2023.\cite{icicle-sw-comps}  Those components can be found in the VC3 release github repository.\cite{rehs22-github-icicle}. Below, we highlight several of our projects.

\subsection{Jupyter Notebook Clients}
\label{sec:notebook-apps}
\noindent We created various Jupyter Notebooks that displayed important components in Authentication and Authorization, as well as setting up services like a Neo4j Knowledge Graph on a Tapis Pod. We explore how to populate blank templates of knowledge graphs, and went on to develop notebooks as clients to these knowledge graphs.

\subsubsection{Tapis Authentication Notebook}
\label{sec:proj-tapis-auth}
\noindent We compiled authentication to Tapis as well as creation and authorization for Tapis Pods in this notebook. It forms the basis for many of our other applications.



\subsubsection{Data Loading Notebook}
\label{sec:proj-neorj-data-load}
\noindent In many cases, there is pre-existing data that must be brought into a knowledge graph. We showed how this can be done in our Data Loading notebook, where we displayed CSV data being loaded into a blank knowledge graph. The main requisite is having a link where the CSV is stored; we used Github for this. Then, we simply pass in the link to the Cypher function
\begin{verbatim}
LOAD CSV WITH HEADERS
\end{verbatim}
and specify columns to node types using the 
\begin{verbatim}
MERGE
\end{verbatim}
keyword. We also did some scripting to automate the creation of relationships from CSV data. 

\begin{figure}[h]
\centering
\includegraphics[width=.96\linewidth]{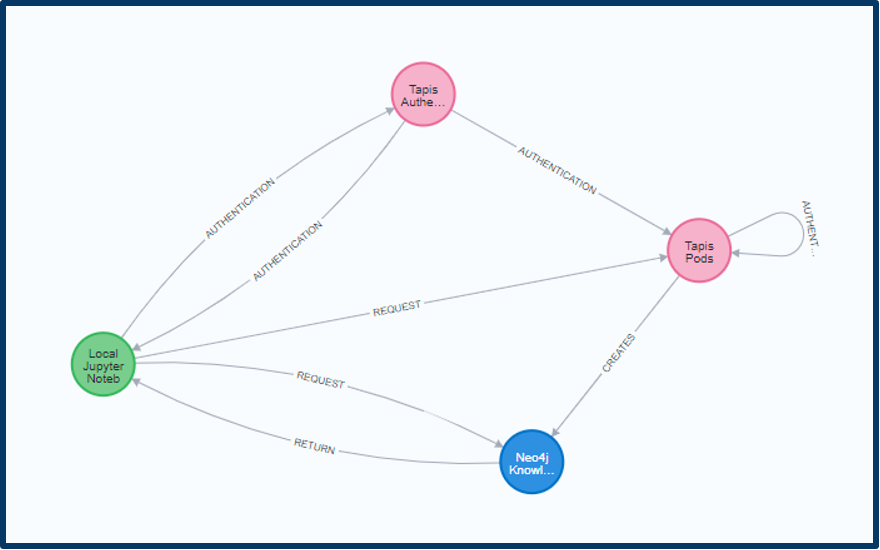}
\caption{Figure kg-graph-visualization.png} 
\label{fig:kg-graph-visualization}
\end{figure}
\subsubsection{Knowledge Graph Visualization of Architecture}
\label{sec:vis-kg-arch}
\noindent This Knowledge Graph is an alternative visualization to our architecture diagram. It contains nodes representing the Tapis Authentication service, the Tapis Pods service, the Neo4j Knowledge Graph, and the Local Jupyter Notebook that queried the knowledge graph. It also contained relationships that displayed how the nodes interacted with each other. The goal of this project was to show how we could successfully host Neo4j Knowledge Graphs on Tapis Pods, and authenticate to them using local Jupyter Notebooks - we created a notebook for this so that the user could experience the point of view from the "Local Jupyter Notebook" node.

Figure~\ref{fig:kg-graph-visualization} shows the different nodes and relationships in the KG.

\subsection{Command-line Interfaces (CLIs) }
\label{sec:results-console-apps}
\noindent We developed Command-line Interfaces (CLIs) which will allow users to connect to HPC systems using a simple terminal/shell application. The first step to any communication with a TACC system (cluster, Tapis service, etc.) is authentication. The prerequisite to this is that one must have an account registered with TACC already; this can be easily checked using the online TACC portal. Our console applications handle authentication to Tapis with the user's TACC account. We demonstrate the authentication process in the “tapis\_auth” notebook described in Section~\ref{sec:proj-tapis-auth}; our CLIs simply transported that functionality to the console. We developed two separate CLIs. One had the goal of streamlining Tapis functions such as creation of pods (TapisCL-ICICLE), while the other had the goal of easily working with data in a Neo4j pod (ICICONSOLE). As other institutions begin using Tapis services, our applications will also be to authenticate to them.

\begin{figure}[h]
\centering
\includegraphics[width=.8\linewidth]{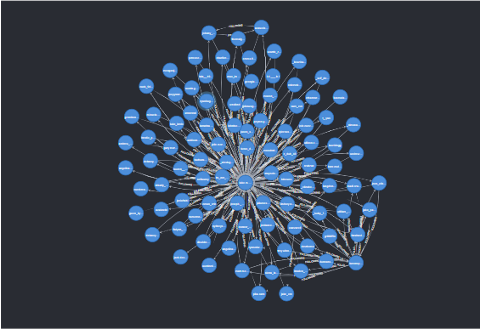}
\caption{Instagram Web Scraper Knowledge Graph.} 
\label{fig:web-scraping}
\end{figure}

\subsection{Application Examples}
\subsubsection{Instagram Web Scraper Knowledge Graph}
\label{sec:proj-web-scrape}
\noindent We also developed an application to gather follower data from Instagram and feed that information into a Neo4j pod hosted on the TACC Tapis service. Since every user's data was stored in a separate JSon file, it offered an opportunity to explore different ways of loading information into Neo4j graphs. After uploading the data to the pod, we were able to visualize and query individual users and their relationships on the Tapis pod shown in Figure~\ref{fig:web-scraping}.

\begin{figure}[h]
\centering
\includegraphics[width=.96\linewidth]{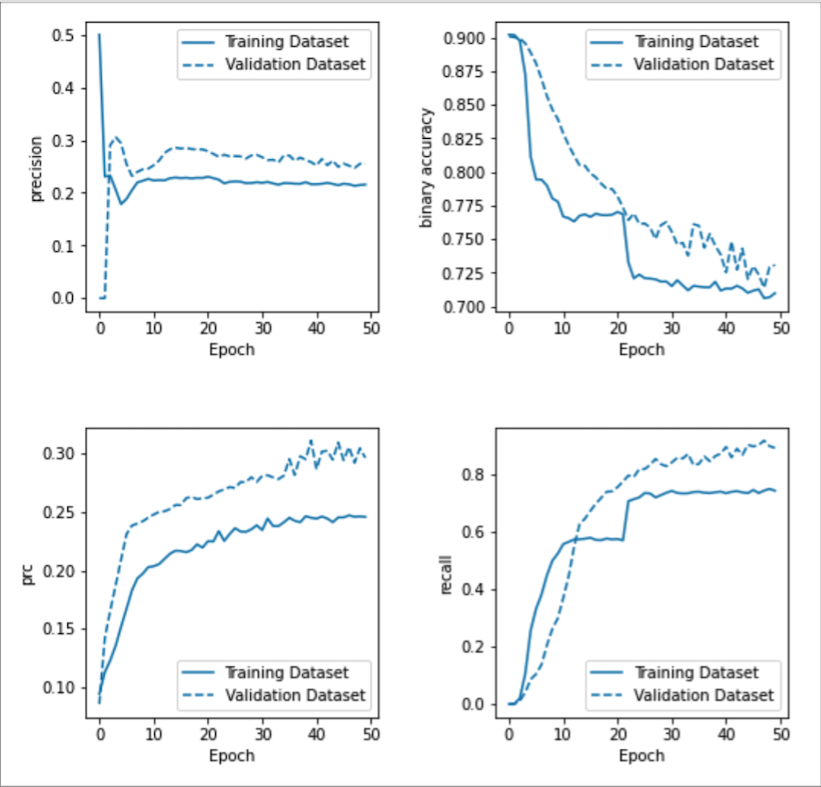}
\caption{Train and test results for the ML Asteroid classification model.} 
\label{fig:train-and-test-data-graphs}
\end{figure}

\subsubsection{Neural Network Asteroid Classification}
\label{sec:proj-aster}
\noindent In order to demonstrate the capabilities and flexibility of Neo4j KG system we developed, we created a dense neural network to classify asteroids as hazardous or harmless using data from NASA’s, “Nearest Earth Object,” dataset.\cite{nasa-neos-dataset} This classification was done based on diameter, absolute magnitude, relative velocity, and estimated miss distance of an asteroid. Because of the data imbalance between hazardous and non-hazardous asteroids in the dataset, we achieved only about 75\% accuracy during dataset testing. However, only 3.8\% of these predictions were false negatives (that is a hazardous asteroid was classified as non-hazardous).

\begin{figure}[h]
\includegraphics[width=.86\linewidth]{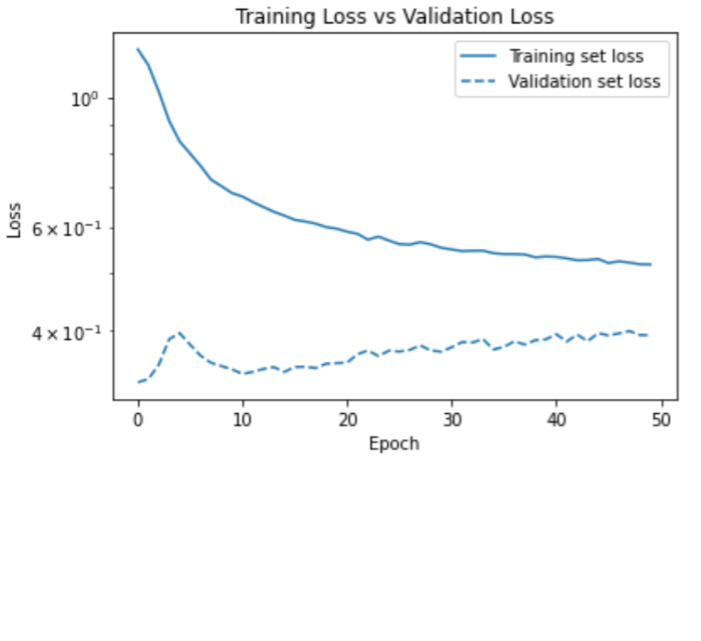}
\squeezeupa
\squeezeupa
\squeezeupa
\squeezeupa
\caption{Training set losses for the ML Asteroid classification model} 
\label{fig:train-set-loss-graphs}
\end{figure}

Figures~\ref{fig:train-and-test-data-graphs} and~\ref{fig:train-set-loss-graphs} visualize the performance metrics of the machine learning model. Note that in Fig~\ref{fig:train-and-test-data-graphs}, PRC represents the area under the precision curve (AUPRC).

\begin{figure}[h]
\centering
\includegraphics[width=.86\linewidth]{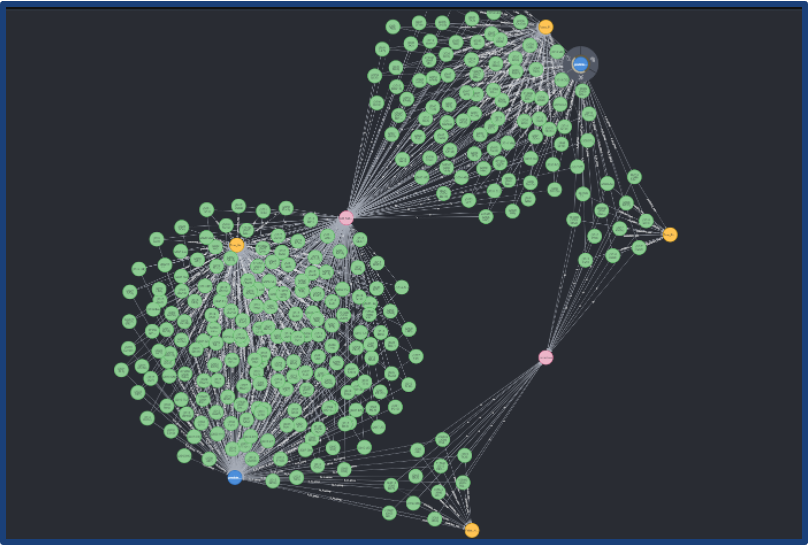}
\caption{Training vs. Testing accuracy with Dropout Regularization.}
\label{fig:asteroid-neural-networks-knowledge-graphs}
\end{figure}

These deficiencies, however, proved valuable, since visualizing the model's faulty predictions gave us more opportunity to push our data loading program. Then we visualized these relationships in Figure~\ref{fig:asteroid-neural-networks-knowledge-graphs}.

In Figure~\ref{fig:asteroid-neural-networks-knowledge-graphs}, green nodes represent asteroids. Blue nodes represent the asteroids’ classifications, as well as how the neural network classified them (hazardous or non-hazardous), and orange nodes represent the nature of the network’s prediction (false negative, false positive, etc.)

While feeding a neural network’s output into a knowledge graph isn’t the most obvious usage of Neo4j, it demonstrates that it can present complex, high volume data in a unique visual medium using simple, short pieces of code; and it can also do so without extensive knowledge of the backend that goes into creating these visualizations. Hosting the knowledge graph on a Tapis pod during this project also demonstrated its efficacy for similar use cases.

\section{Conclusions and Future Work}
\label{sec:conclusions}
\noindent Throughout the course of our work, our team developed notebooks for stable, efficient authentication to ICICLE services hosted on TACC using Tapis, notebooks to host Neo4j Knowledge graphs, and command line applications to make accessing HPC systems more user-friendly. Through our research, we created the first instances of Jupyter Notebooks to connect to the iTapis/ICICLE system and access ICICLE-hosted KGs behind a Kubernetes Pods. Our interactive Jupyter Notebook Clients successfully address the issue of democratizing HPC systems by providing a simple and accessible way to create remotely hosted knowledge graphs by providing an easy plug-and-play that simplifies the authentication process. A key outcome of our efforts includes contributing our software to the upcoming ICICLE Software Components Catalog, which will be released around April, 2023.\cite{rehs22-github-icicle, icicle-sw-comps} 

Going forward, we will work on expanding the capabilities of our applications to ensure greater outreach of HPC resources. One of our primary goals is to develop our console applications to a production standard for easy access to Tapis resources. Additionally, we want to work with Visual Analytics within ICICLE to visualize our data outside of Neo4j.

\section{Acknowledgment}
\label{sec:acknowledgement}
\noindent As members of REHS 2022 Program, we are grateful to have been a part of such a large and impactful project, and we hope that the materials presented in this paper will be of use for ICICLE developers. The work on this project was carried out with the support of the San Diego Supercomputer Center Research Experience for Undergraduates (REHS) program. We also want to acknowledge the use of several NSF funded resources and services including: the AI Institute for Intelligent CyberInfrastructure with Computational Learning in the Environment (ICICLE) (\#2112606); SDSC Expanse project (\#1928224); TACC Stampede System (\# 1663578) and Tapis projects (\#1931439); the NSF Track 3 Award: COre National Ecosystem for CyberinfrasTructure (CONECT) (\#2138307); and finally, the Extreme Science and Engineering Discovery Environment (XSEDE) (NSF award \#ACI-1548562).



%

\bibliographystyle{IEEEtran}
\bibliography{rehs-2022-refs}


\end{document}